# Electric-Field-Induced Coherent Control of Nitrogen Vacancy Centers

Gerald Q. Yan[1], Senlei Li[1], Tatsuya Yamamoto[2], Mengqi Huang[1], Nathan McLaughlin[1], Takayuki Nozaki[2], Hailong Wang[3], Shinji Yuasa[2], and Chunhui Rita Du[1,3]

[1]Department of Physics, University of California, San Diego, La Jolla, California 92093, USA
[2]Research Center for Emerging Computing Technologies, National Institute of Advanced Industrial Science and Technology (AIST), Tsukuba, Ibaraki, 305-8568, Japan
[3]Center for Memory and Recording Research, University of California, San Diego, La Jolla, California 92093-0401, USA

Enabling scalable and energy-efficient control of spin defects in solid-state media is desirable for realizing transformative quantum information technologies. Exploiting voltage-controlled magnetic anisotropy, we report coherent manipulation of nitrogen-vacancy (NV) centers by the spatially confined magnetic stray fields produced by a proximate resonant magnetic tunnel junction (MTJ). Remarkably, the coherent coupling between NV centers and the MTJ can be systematically controlled by a DC bias voltage, allowing for appreciable electrical tunability in the presented hybrid system. In comparison with current state-of-the-art techniques, the demonstrated NV-based quantum operational platform exhibits significant advantages in scalability, device compatibility, and energy-efficiency, further expanding the role of NV centers in a broad range of quantum computing, sensing, and communications applications.

Over the past decade, nitrogen-vacancy (NV) centers, optically active spin defects in diamond with single-spin addressability, excellent quantum coherence, and remarkable functionality over a broad range of temperatures, have emerged as a promising platform for developing transformative quantum information sciences and technologies [1–3]. NV centers have already been successfully applied to quantum sensing [2–4], imaging [5–8], communication [9–11], and network research [10–12], enabling unprecedented field sensitivity, nanoscale spatial resolution, and long-range qubit transmission. Hybrid systems establishing strong coupling between NV centers, photons, and other solid-state media for applications in functional quantum devices are also currently being developed [13–30].

Despite significant progress, problems related to decoherence, scalability, and control of entanglement remain to be solved to fully realize the potential of NV centers for developing novel quantum processors [16,30–33]. In the current state-of-the-art, the quantum spin state of NV centers is typically manipulated by spatially dispersive Oersted fields generated by radiofrequency (RF) electric currents [34,35], imposing an inherent challenge for achieving highly local and scalable manipulation of NV centers. In addition, the current-induced Joule heating inherent to this approach can generate substantial thermal noise, resulting in decoherence of the NV centers [36]. These long-standing issues have thus far hindered the development of NV-based quantum computing platforms and proven profoundly difficult to overcome.

To address this challenge, we integrate NV centers with a functional magnetic tunnel junction (MTJ) device exhibiting voltage-controlled magnetic anisotropy (VCMA) [37–39], enabling electric-field-driven coherent control of NV centers in an energy-efficient manner. Manipulation of the spin qubits is achieved by exploiting the spatially confined magnetic stray fields produced by a resonant magnetic free layer within the MTJ stack. It is worth noting that the millisecond-long NV spin relaxation times are preserved in the presented hybrid system, offering an attractive platform for developing high-density, scalable, NV-based, solid-state architectures for applications at the forefront of quantum sciences and technologies.

We first describe the detailed structure of the device used in our measurements, as illustrated in Figure 1(a). An MTJ composed of (from bottom to top): Ta(5 nm)/$Co_{40}Fe_{40}B_{20}$(1 nm)/MgO(2 nm)/$Co_{56}Fe_{24}B_{20}$(5 nm)/Ru(7 nm)/Cr(5 nm)/Au(50 nm) was fabricated on a Si substrate (see Supplementary Information for details). The in-plane magnetized $Co_{56}Fe_{24}B_{20}$ serves as a fixed reference layer while $Co_{40}Fe_{40}B_{20}$ forms the free layer with spontaneous perpendicular magnetization due to a weak out-of-plane anisotropy [40–44]. Viewed from above, the MTJ has an elongated hexagonal shape with a length of 6 μm and a width of 2 μm. A diamond microchip [7] with lateral dimensions of 30 μm × 30 μm was placed on top of the MTJ. NV centers were shallowly implanted on the bottom surface of the diamond chip. An optical image shown in Figure 1(b) provides an overview of a prepared device. In our experiments, NV centers were optically addressed using either confocal or wide-field microscopy [6,45–47]. To enable coherent control of the NV centers, we take advantage of the oscillating magnetic stray fields generated by the proximate resonant MTJ. Applying a voltage across the MTJ modifies the electron charge or spin densities at the $Co_{40}Fe_{40}B_{20}$/MgO interface and induces a variation of the magnetic anisotropy in the free layer through the spin-orbit interaction [39,48–51]. Due to the VCMA, the magnetic easy axis varies between the out-of-plane and in-plane directions depending on the sign and magnitude of the applied voltage. Therefore, applying an RF voltage to the MTJ can excite coherent magnetic oscillations of the $Co_{40}Fe_{40}B_{20}$ free layer under an appropriate, static external magnetic field.

The experimental measurements of VCMA-driven ferromagnetic resonance (FMR) of the MTJ using a homodyne detection technique [37] are shown in Figure 1(a). An external magnetic field $B_{ext}$ is applied at an angle of 54 degrees relative to the out-of-plane direction, in alignment with the corresponding NV axis. The in-plane projection of $B_{ext}$ lies along the long-axis of the MTJ, and an RF voltage with a power of −2 dBm is applied across the MTJ via the RF input port of a bias tee. Under FMR conditions, the oscillating tunnel magnetoresistance of the MTJ couples with the RF tunnel current to produce a significant DC voltage $V$, which can be detected through the DC port of the bias tee. In this way, an output DC voltage was measured as a function of input frequency. Figure 1(c) plots a two-dimensional (2D) map of the derivative of the measured DC voltage with respect to frequency as a function of $B_{ext}$ and frequency $f$ of the applied RF voltage. VCMA-induced FMR is observed with expected field and frequency dependence, in accordance with theoretical predictions (see Supplementary Information for details) . The FMR dispersion curve intersects with the NV electron spin resonance (ESR) frequency at a field $B_{ext}$ = 447 G and frequency $f$ = 1.62 GHz. Under a fixed external magnetic field $B_{ext}$ = 490 G., the resonant frequency of the MTJ can be tuned with VCMA by varying the DC bias voltage across the device, as shown in Figure 1(d).

Next, we utilize confocal microscopy (see Supplementary Information for details) to perform Rabi oscillation measurements of NV centers located in a ~1 μm × 1 μm area of the diamond that is directly above the MTJ, demonstrating electric-field-driven coherent control of NV centers. Before discussing the experimental details, we first briefly review the pertinent physical properties of NV centers. An NV center consists of a substitutional nitrogen atom adjacent to a carbon atom vacancy in one of the nearest neighboring sites of a diamond crystal lattice [1–3]. The negatively charged NV state has an $S$ = 1 electron spin and serves as a three-level qubit system. When an oscillating magnetic field at the NV ESR frequencies $f_{\pm}$ is applied at the NV site, the NV occupation probabilities will periodically oscillate between two different spin levels. These are referred to as Rabi oscillations [1–3], and are illustrated in Figure 2(a). Here, $f_{\pm}$ denote the NV ESR frequencies corresponding to the spin transition between the $m_s$ = 0 and $m_s$ = ±1 states. The top panel of Figure 2(a) shows the measurement protocol used in the Rabi oscillation measurements. A 3-μs-long green laser pulse is first applied to initialize the NV centers to the $m_s$ = 0 state. Next, an RF voltage pulse at the NV ESR frequencies is applied to excite FMR of the MTJ. When the resonant frequency $f_{FMR}$ matches $f_{\pm}$, oscillating stray fields generated by the MTJ will enhance the rate of the $m_s$ = 0 ↔ ±1 NV spin transitions. Compared with the $m_s$ = 0 state, the optically excited $m_s$ = ±1 states are more likely to decay through a non-radiative intersystem crossing before relaxing back to the $m_s$ = 0 ground state, exhibiting reduced photoluminescence (PL) [1,2]. Lastly, a second green laser pulse is applied to readout the NV spin state via the spin-dependent PL. The duration $t$ of the RF voltage pulse is systematically varied to probe the time-dependent variation of the NV PL. A DC bias voltage is also applied during the above measurements, providing electrical tunability of the coherent NV-magnet coupling.

Figure 2(b) shows measured NV Rabi oscillation spectra as a function of $t$ at different NV ESR frequencies. When $f_-$ is detuned from the resonant frequency $f_{FMR}$ of the MTJ such that $|f_- - f_{FMR}| \geq 0.3$ GHz, the measured PL spectrum is essentially independent of the duration of the RF voltage pulse, indicating negligible NV-magnet coupling. As $f_-$ approaches $f_{FMR}$, the measured PL spectra exhibit characteristic periodic oscillations with a Rabi frequency $f_{Rabi}$ of approximately ~5 MHz at a detuning of ±0.1 GHz. Remarkably, when $f_- = f_{FMR}$, we observe significantly faster oscillatory behavior of the NV PL spectrum with an enhanced $f_{Rabi}$ of ~10 MHz, demonstrating robust, electric-field-induced coherent control of NV spin states. The dramatic enhancement of the NV spin rotation rate is driven by the phase synchronization between the resonant MTJ and proximate NV centers. The Rabi oscillation frequency is proportional to the magnitude of the local magnetic stray field $B_\perp$ transverse to the NV spin orientation at the NV site, and is calculated to be 4.8 G, in agreement with theoretical calculations (see Supplementary Information for details). By applying a DC bias voltage $V_B$ to vary the resonant frequency $f_{FMR}$, we are able to further achieve effective tuning of the NV Rabi oscillation frequency through electric-field induced variation of the static magnetic anisotropy [52,53], as shown in Figure 2(c). Figure 2(d) plots the measured $f_{Rabi}$ as a function of $f_-$ for three different DC bias voltages. $f_{FMR}$ shifts towards higher (lower) frequencies with positive (negative) $V_B$, due to a decrease (increase) of the perpendicular magnetic anisotropy. Peak values of $f_{Rabi}$ are consistently observed when $f_-$ matches $f_{FMR}$, demonstrating that the coherent NV spin rotation is indeed driven by VCMA-induced FMR of the MTJ.

We next utilize wide-field magnetometry (see Supplementary Information for details) to further illustrate the presented NV control scheme and its electrical tunability. The laser beam spot width used in the wide-field measurements is approximately ~30 µm × 30 µm, allowing simultaneous imaging of all the NV centers in the diamond microchip that are positioned above the MTJ device. The NV fluorescence is captured using a CMOS camera. Figures 3(a)-3(e) show a representative series of 2D widefield images of the measured NV Rabi oscillation rate at five detuning frequencies ($f_- - f_{FMR}$) corresponding to points "A" to "E" shown in Figure 3(f). At $f_-$ = 1.49 GHz [Figure 3(a)], the measured NV Rabi oscillation frequency $f_{Rabi}$ is nearly negligible over the entire measured area due to a significant mismatch between the NV ESR frequency and MTJ resonant frequency. Moderate NV spin rotation emerges when $f_-$ = 1.57 GHz [Figure 3(b)], suggesting establishment of dipole-mediated coherent coupling between the resonant MTJ and NV centers. The coupling strength is maximized when $f_-$ matches $f_{FMR}$, leading to the highest Rabi oscillation frequencies of the NV centers directly above the MTJ [Figure 3(c)]. Further increasing $f_-$ results in detuning of the NV center and VCMA FMR frequencies and subsequent suppression of the Rabi oscillation frequencies, as illustrated in Figure 3(d) and 3(e). Note that the inhomogeneous distribution of the Rabi frequencies over the device indicates a spatially nonuniform distribution of the magnetic stray field magnitudes produced by the MTJ, which could be tied to underlying magnetic inhomogeneities and domains in the device.

Next, we consider the potential advantages of the presented NV-based quantum operational platform. In contrast with conventional NV-control schemes using spatially dispersive Oersted fields generated by RF currents, the NV-MTJ hybrid device utilizes magnetic stray fields, which are more spatially confined due to dipole-dipole interactions, to enable coherent control of NV centers. To better illustrate this point, Figure 3(g) shows one-dimensional profiles of the local magnetic stray field $B_\perp$ (transverse to the NV axis) measured across the short-axis of the MTJ at three NV ESR frequencies. Notably, $B_\perp$ shows a finite value at positions within the width of the MTJ device. At positions beyond the width of the device, $B_\perp$ quickly decays to a vanishingly small value on a length scale of ~100 nm. Our experimental results agree with theoretical calculations (see Supplementary Information for details), confirming the highly localized NV control strategy demonstrated by the presented NV-MTJ device. This merit is particularly beneficial to the development of high-density NV-based information processing and storage technologies, where minimal crosstalk between neighboring operational units is desirable [19,54,55]. The solid-state nature of the MTJ devices and NV centers renders them readily compatible with a large family of functional quantum architectures, promoting the use of NV centers in implementing large-scale integrated quantum networks [11]. The voltage-controlled coherent NV-magnet coupling is further illustrated in Figure 4(a)-4(c). At $B_{ext}$ = 460 G and $f_-$ = 1.58 GHz, application of a negative bias voltage of −0.4 V significantly enhances the coherent NV rotation rate [Figure 4(a)], whilst a positive voltage effectively suppresses the NV Rabi frequencies [Figure 4(c)]. Our results demonstrate that the coherent NV-magnet coupling can be electrically switched on and off by a moderate bias voltage. We expect that further optimization of the material/device parameters could add appreciable tunability of the NV-MTJ hybrid device across a broad range of experimental conditions.

Lastly, we perform NV spin relaxometry measurements using confocal microscopy to ascertain the spin relaxation times ($T_1$) of the NV centers directly above the MTJ. Our measurements were performed at $B_{ext}$ = 438 G and $f_-$ = 1.64 GHz. At $V_B$ = 0 V, $f_-$ is close to $f_{FMR}$, which is the minimum frequency of the magnon band. Application of a DC bias voltage shifts $f_{FMR}$ higher or lower than $f_-$ depending on the sign of the voltage, as illustrated in Figure 4(d). The top panel of Figure 4(e) shows the optical and RF sequence used in the NV relaxometry measurements. A green laser pulse was first applied to initialize the NV spins to the $m_s$ = 0 state. After a delay time $t$, we measured the occupation probabilities of the $m_s$ = −1 states by applying an RF $\pi$ voltage pulse at the corresponding ESR frequencies and measuring the spin-dependent PL with a green-laser readout pulse. Note that the RF pulses were delivered with an RF voltage applied across the MTJ. The bottom panel of Figure 4(e) plots the measured NV PL as a function of delay time $t$ at three different bias voltages. By fitting the results with an exponential decay function: $PL(t) = A_0 + A e^{-t/T_1}$ [35,56], where $A$ and $A_0$ are constants, the $T_1$ time of the NV centers is measured to be 1429 μs, 588 μs, and 175 μs for bias voltages $V_B$ of +0.3 V, 0 V, and −0.3 V, respectively. The electrical tunability of $T_1$ exploits the voltage-controlled $f_{FMR}$ of the magnetic free layer, which determines the magnitude of spin noise at the NV ESR frequency $f_-$ and leads to relaxation of the NV center spins [35]. Note that the $T_1$ of NV centers with $f_-$ tuned slightly above the magnon band ($V_B$ > 0 V) is comparable to the intrinsic relaxation time of NV centers positioned away from the MTJ ($T_1$ = 1667 μs), which is slightly shorter than the value reported in Ref. 36. In the future, it would also be valuable to characterize the coherence time of NV centers in the hybrid system.

In summary, we have demonstrated coherent control of NV centers by a resonant MTJ. Exploiting VCMA-driven oscillating stray fields, we achieved local control and electrically tunable control of NV centers, offering new opportunities for developing scalable, high-density NV-based quantum operational units. Due to the large electrical resistance of the insulating tunnel layer of the device, the magnitude of the RF electric current flowing through the MTJ is in the microampere regime, virtually eliminating ohmic losses and the resulting thermal decoherence of nearby NV centers. We remark that the millisecond long spin relaxation time remains preserved in NV centers positioned in nanoscale proximity to the MTJ. These merits highlight the presented hybrid solid-state system as a promising building block for a broad range of transformative applications in NV-based quantum computing, sensing, and networking [5,11,16,28,54,55,57,58].

**Acknowledgements.**
The authors would like to thank Hanyi Lu for assistance in sample preparation. The authors thank Yaroslav Tserkovnyak, Ilya N. Krivorotov, Pramey Upadhyaya, and Avinash Rustagi for insightful discussions. G. Q. Y., S. L., M. H., and C. R. D. were supported by U. S. National Science Foundation (NSF) under award ECCS-2029558 and DMR-2046227. N. J. M., H. W., and C. R. D. acknowledged the support from the Air Force Office of Scientific Research under award FA9550-20-1-0319 and its Young Investigator Program under award FA9550-21-1-0125.

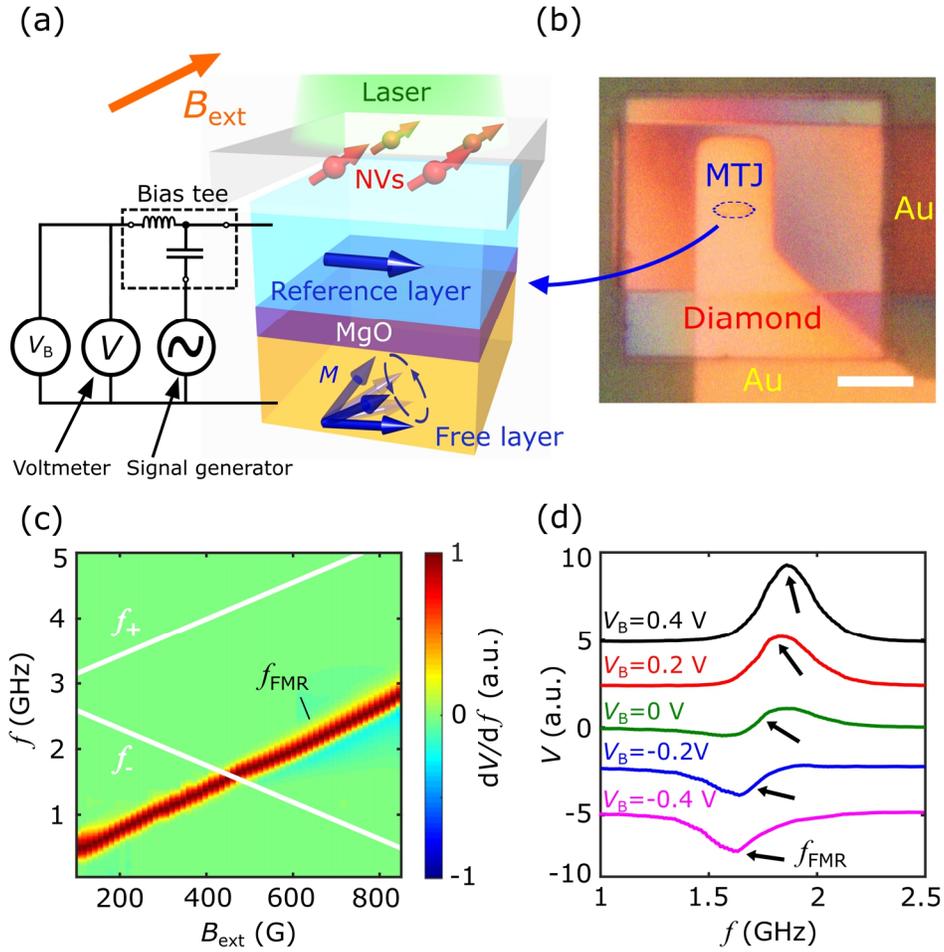

**Figure 1.** (a) Schematic illustration of a nitrogen-vacancy (NV)-magnetic-tunnel-junction (MTJ) hybrid device. A diamond microchip containing NV ensembles is placed on top of a prepared MTJ. The NV spin state is optically addressed using confocal or wide-field microscopy. Electrical excitation and detection of VCMA-induced magnetic resonance utilizes a standard homodyne detection circuit. (b) Optical image showing a prepared NV-MTJ device. The blue dashed line outlines the edges of the MTJ, and the scale bar is 10 μm. (c) Derivative of measured DC voltage $dV/df$ as a function of external magnetic field $B_{ext}$ and frequency $f$ of the applied voltage. The white lines represent the field-dependent upper and lower NV electron spin resonance frequencies $f_{\pm}$. (d) DC bias voltage induced variation of spectrum shape and shift of the resonant frequency $f_{FMR}$ of a MTJ device measured at $B_{ext}$ = 490 G.

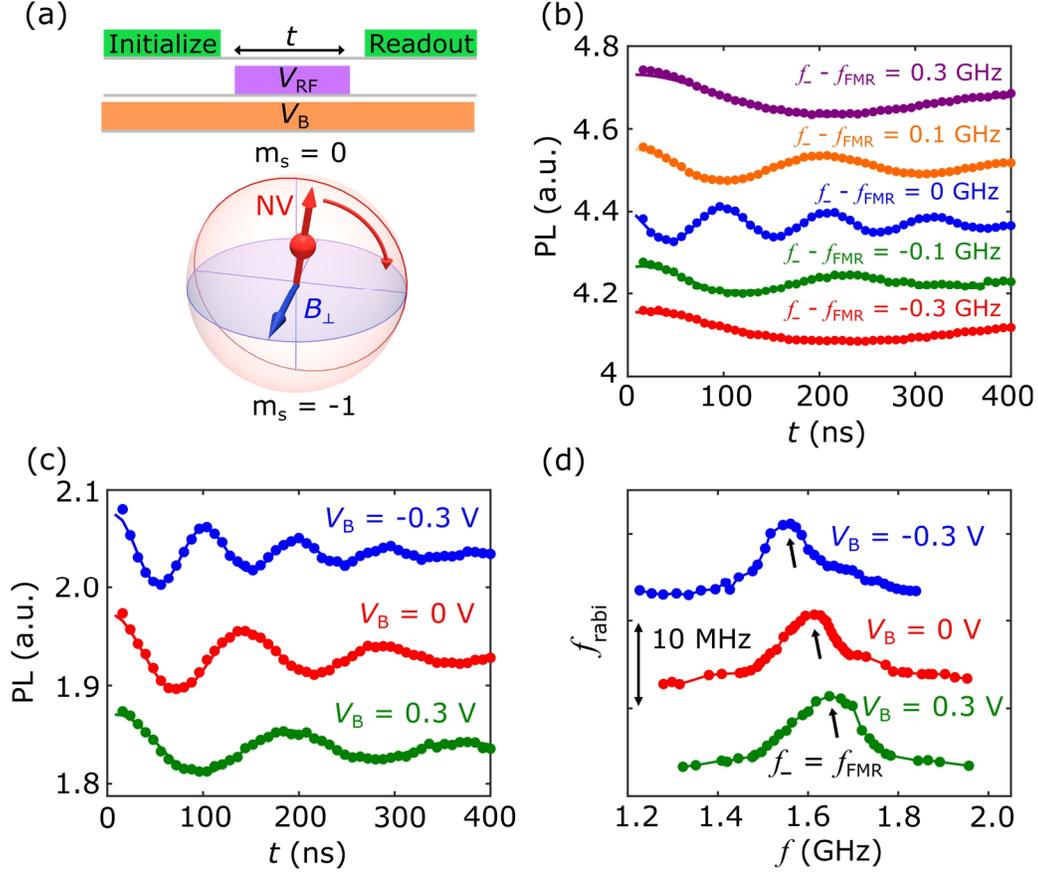

**Figure 2.** (a) Top panel: Optical and RF voltage pulse sequence used in the NV Rabi oscillation measurements. Bottom panel: Schematic of NV Rabi oscillations between the $m_s = 0$ and $m_s = -1$ states on the Bloch sphere. (b) NV photoluminescence (PL) intensity as a function of the delay time $t$ measured at detuning frequencies ($f_- - f_{FMR}$) of $\pm 0.3$ GHz, $\pm 0.1$ GHz, and $0$ GHz. (c) Time dependent NV PL spectra measured with the application of DC bias voltages $V_B$ of $-0.3$ V, 0 V, and 0.3 V under an external magnetic field $B_{ext} = 470$ G. The Rabi oscillation frequency is enhanced or suppressed depending on the sign of the DC bias voltage. (d) Measured Rabi frequency as a function of NV electron spin resonance (ESR) frequency $f_-$ under three different DC bias voltages. The curves are vertically offset for visual clarity.

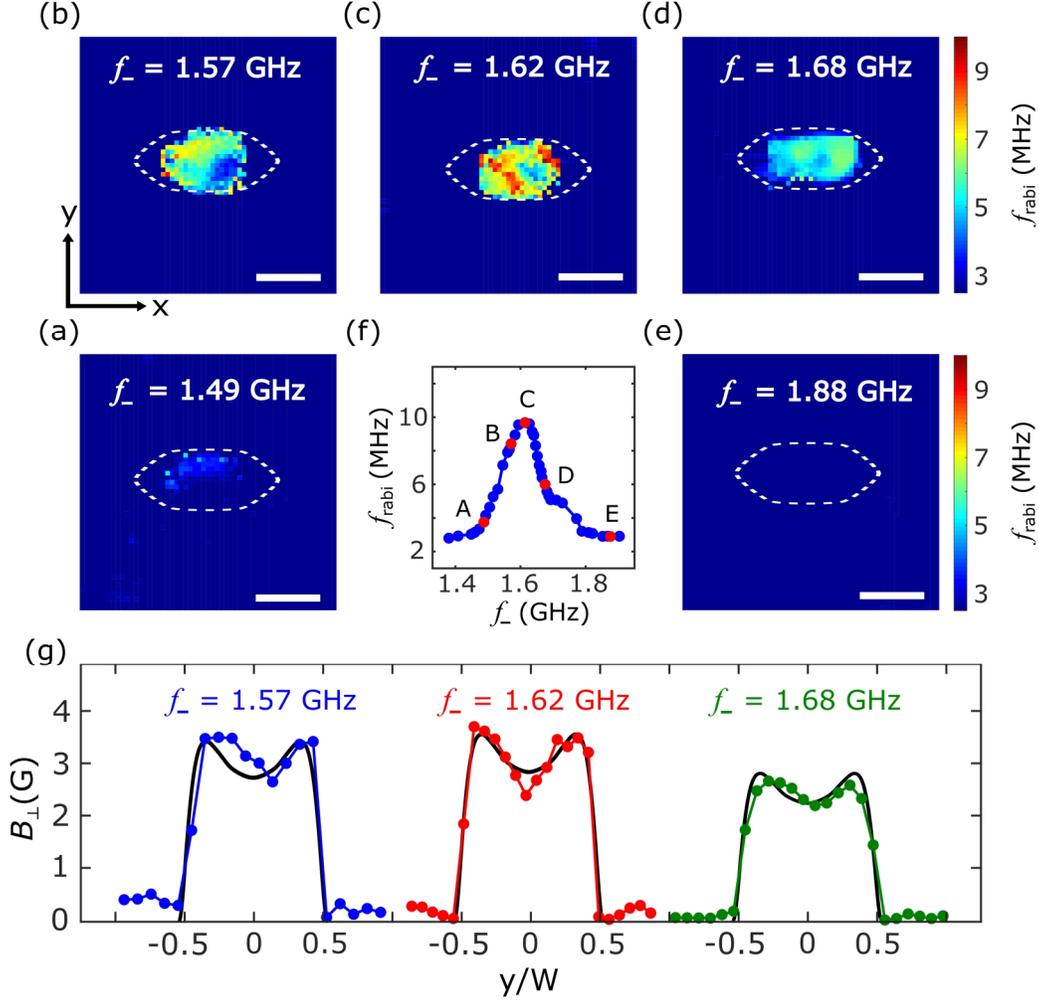

**Figure 3**. (a)-(e) Two-dimensional (2D) maps of the Rabi oscillation frequency measured at NV ESR frequencies $f_-$ of 1.49, 1.57, 1.62, 1.68, and 1.88 GHz, respectively. The white dashed line outlines the lateral boundary of the MTJ underneath the diamond chip, and the scale bar is 2 μm. (f) Rabi oscillation frequency $f_{Rabi}$ measured as a function of the ESR frequency $f_-$ for NV centers directly above the center of the MTJ. The points "A" to "E" marked on the curve correspond to the five NV ESR frequencies $f_-$ used in the NV wide-field magnetometry measurements presented in Figs. 3(a)-(e). (g) Linecuts of extracted magnetic stray field $B_\perp$ (transverse to the NV axis) along the short axis ($y$ axis) of the MTJ at ESR frequencies of 1.57, 1.62, and 1.68 GHz. Micromagnetic simulation results (black curves) are in qualitative agreement with the experimental data. The unit of the $y$ axis has been normalized by the short-axis width ($W$ = 2 μm) of the MTJ device.

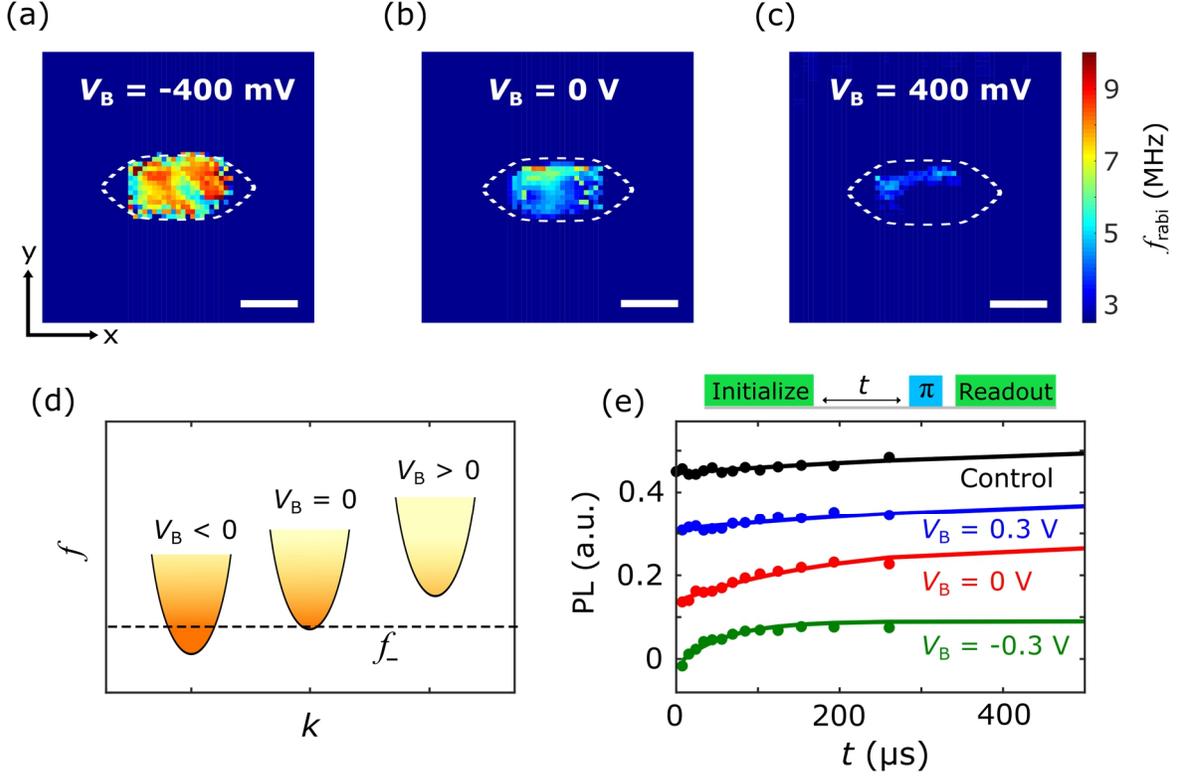

**Figure 4**. (a)-(c) 2D maps of Rabi oscillation frequency $f_{Rabi}$ measured at $f_- = 1.58$ GHz with a bias voltage of $-0.4$ V (a), 0 V (b), and $+0.4$ V (c), respectively, under an external magnetic field $B_{ext} = 460$ G. The white dashed line outlines the lateral boundary of the MTJ, and the scale bar is 2 μm. (d) Schematic showing the magnon dispersions for various bias voltages and their intersection with the NV ESR frequency $f_-$. The magnon occupation of the magnetic free layer follows the Bose-Einstein distribution as indicated by the fading color intensities. A positive (negative) bias shifts the magnon band upwards (downwards) relative to $f_-$. The dispersions have been shifted in $k$ for visual clarity. (e) Top panel: Optical and RF voltage pulse sequence used in the NV relaxometry measurements. Bottom panel: NV PL intensity measured as a function of the delay time $t$, from which the NV relaxation time $T_1$ is extracted to be 1429 μs, 588 μs, and 175 μs under the application of DC bias voltages of 0.3 V, 0 V, and $-0.3$ V, respectively. Control measurements of NV centers in the diamond microchip give a $T_1$ time of 1667 μs.